\documentclass{pasj00}
\newcommand{\fermi}{{\it Fermi}-LAT}
\newcommand{\src}{1H~0323+342}
\newcommand{\oister}{{\it OISTER}}
\newcommand{\kanata}{{\it Kanata}}

\usepackage{natbib}

\begin{document}
\SetRunningHead{Itoh et al.}{{\it Kanata} and {\it OISTER} observations of narrow line Seyfert 1 galaxy 1H~0323+342}
%\Received{}%{yyyy/mm/dd}
\Accepted{2014/08/03}%{yyyy/mm/dd}
%\Published{}%{yyyy/mm/dd}

\title{Variable optical polarization during high state in $\gamma$-ray loud narrow line Seyfert 1 galaxy 1H~0323+342}

%%% begin:list of authors
% Do NOT capitalize all letters in "textsc".
%\author{A-Firstname \textsc{A-Familyname} %
 % \thanks{Example: Present Address is xxxxxxxxxx}}
%\affil{A-Address of Institute}
%\email{aaaaa@xxx.xxx.xx.xx}

%\author{B-Firstname \textsc{B-Familyname}}
%\affil{B-Address of Institute}\email{bbbbb@xxx.xxx.xx.xx}
%\and
%\author{C-Firstname {\sc C-Familyname}}
%\affil{C-Address of Institute}\email{ccccc@xxx.xxx.xx.xx}
%%% end:list of authors

%%% Please use the following style in case that sorting by 
%%% affiliation is impossible. 
%
\author{
Ryosuke \textsc{Itoh}\altaffilmark{1},
Yasuyuki~T. \textsc{Tanaka}\altaffilmark{2},
Hiroshi \textsc{Akitaya}\altaffilmark{2},
Makoto \textsc{Uemura}\altaffilmark{2},
Yasushi \textsc{Fukazawa}\altaffilmark{1},
Yoshiyuki \textsc{Inoue}\altaffilmark{3},
Akihiro \textsc{Doi}\altaffilmark{3},
Akira \textsc{Arai}\altaffilmark{4},
Hidekazu \textsc{Hanayama}\altaffilmark{5},
Osamu \textsc{Hashimoto}\altaffilmark{6},
Masahiko \textsc{Hayashi}\altaffilmark{7},
Hideyuki \textsc{Izumiura}\altaffilmark{8},
Yuka \textsc{Kanda}\altaffilmark{1},
Koji~S. \textsc{Kawabata}\altaffilmark{2},
Kenji \textsc{Kawaguchi}\altaffilmark{1},
Nobuyuki \textsc{Kawai}\altaffilmark{9},
Kenzo \textsc{Kinugasa}\altaffilmark{10},
Daisuke \textsc{Kuroda}\altaffilmark{8},
Takeshi \textsc{Miyaji}\altaffilmark{5},
Yuki \textsc{Moritani}\altaffilmark{2},
Tomoki \textsc{Morokuma}\altaffilmark{11},
Katsuhiro~L. \textsc{Murata}\altaffilmark{12},
Takahiro \textsc{Nagayama}\altaffilmark{13},
Yumiko \textsc{Oasa}\altaffilmark{14},
Tomohito \textsc{Ohshima}\altaffilmark{15},
Takashi \textsc{Ohsugi}\altaffilmark{2},
Yoshihiko \textsc{Saito}\altaffilmark{9},
Shuuichiro \textsc{Sakata}\altaffilmark{13},
Mahito \textsc{Sasada}\altaffilmark{15},
Kazuhiro \textsc{Sekiguchi}\altaffilmark{7},
Yuhei \textsc{Takagi}\altaffilmark{4},
Jun \textsc{Takahashi}\altaffilmark{4},
Katsutoshi \textsc{Takaki}\altaffilmark{1},
Takahiro \textsc{Ui}\altaffilmark{1},
Makoto \textsc{Watanabe}\altaffilmark{16,17},
Masayuki \textsc{Yamanaka}\altaffilmark{18},
Satoshi \textsc{Yamashita}\altaffilmark{13} and
Michitoshi \textsc{Yoshida}\altaffilmark{2}
}

\altaffiltext{1}{Department of Physical Sciences, Hiroshima
University, Higashi-Hiroshima, Hiroshima 739-8526, Japan}
\email{itoh@hep01.hepl.hiroshima-u.ac.jp}
\altaffiltext{2}{Hiroshima Astrophysical Science Center, Hiroshima
University, Higashi-Hiroshima, Hiroshima 739-8526, Japan}
\email{ytanaka@hep01.hepl.hiroshima-u.ac.jp}
\altaffiltext{3}{Institute of Space and Astronautical Science (ISAS)
Japan Aerospace Exploration Agency (JAXA), Sagamihara, Kanagawa
252-5210, Japan}
\altaffiltext{4}{Nishi-Harima Astronomical Observatory, Center for
Astronomy, University of Hyogo, 407-2, Nishigaichi, Sayo-cho, Sayo,
Hyogo 679-5313, Japan}
\altaffiltext{5}{Ishigakijima Astronomical Observatory, National
Astronomical Observatory of Japan, 1024-1 Arakawa, Ishigaki, Okinawa
907-0024, Japan}
\altaffiltext{6}{Gunma Astronomical Observatory, 6860-86 Nakayama,
Takayama, Agatsuma, Gunma 377-0702, Japan}
\altaffiltext{7}{National Astronomical Observatory of Japan, Osawa
2-21-2, Mitaka, Tokyo, 181-8588, Japan}
\altaffiltext{8}{Okayama Astrophysical Observatory, National
Astronomical Observatory of Japan, Honjo 3037-5, Kamogata, Asakuchi,
Okayama 719-0232, Japan}
\altaffiltext{9}{Department of Physics, Tokyo Institute of Technology,
2-12-1 Ookayama, Meguro-ku, Tokyo 152-8551, Japan}
\altaffiltext{10}{Nobeyama Radio Observatory, National Astronomical Observatory of Japan, 
462-2 Nobeyama, Minamimaki,Minamisaku, Nagano 384-1305, Japan}
\altaffiltext{11}{Institute of Astronomy, Graduate School of Science,
The University of Tokyo, 2-21-1 Osawa, Mitaka, Tokyo 181-0015, Japan}
\altaffiltext{12}{Department of Astrophysics, Nagoya University,
Chikusa-ku Nagoya 464-8602, Japan}
\altaffiltext{13}{Graduate School of Science and Engineering,
Kagoshima University, 1-21-35 Korimoto, Kagoshima 890-0065, Japan}
\altaffiltext{14}{Faculty of Education, Saitama University, 255
Shimo-Okubo, Sakura, Saitama, 338-8570, Japan}
\altaffiltext{15}{Department of Astronomy, Graduate School of Science,
Kyoto University, Sakyo-ku, Kyoto 606-8502, Japan}
\altaffiltext{16}{Department of Cosmosciences, Graduate School of
Science, Hokkaido University, Kita-ku, Sapporo 060-0810, Japan}
\altaffiltext{17}{Department of Earth and Planetary Sciences, School
of Science, Hokkaido University, Kita-ku, Sapporo 060-0810, Japan}
\altaffiltext{18}{Kwasan Observatory, Kyoto University, Ohmine-cho
Kita Kazan, Yamashina-ku, Kyoto 607-8471, Japan}

%% `\KeyWords{}' always has to be placed before `\maketitle'.
\KeyWords{galaxies: active; galaxies: jets; galaxies: individual (1H~0323+342); galaxies: Seyfert; radiation mechanisms: non-thermal} %Do NOT move this preamble from here!

\maketitle

\begin{abstract}
We present results of optical polarimetric and multi-band photometric observations for $\gamma$-ray loud narrow-line Seyfert 1 galaxy \src. This object has been monitored by 1.5\,m {\it Kanata} telescope since 2012 September but following a $\gamma$-ray flux enhancement detected by \fermi\ on MJD 56483 (2013 July 10) dense follow-up was performed by ten 0.5--2.0\,m telescopes in Japan over one week. The 2-year $R_{\rm C}$-band light curve showed clear brightening corresponding to the $\gamma$-ray flux increase and then decayed gradually. The high state as a whole lasted for $\sim$20 days, during which we clearly detected optical polarization from this object. The polarization degree (PD) of the source increased from 0--1\% in quiescence to $\sim 3$\% at maximum and then declined to the quiescent level, with the duration of the enhancement of less than 10 days. The moderate PD around the peak allowed us to precisely measure the daily polarization angle (PA). As a result, we found that the daily PAs were almost constant and aligned to the jet axis, suggesting that the magnetic field direction at the emission region is transverse to the jet. This implies either a presence of helical/toroidal magnetic field or transverse magnetic field compressed by shock(s). We also found small-amplitude intra-night variability during the 2-hour continuous exposure on a single night. We discuss these findings based on the turbulent multi-zone model recently advocated by \citet{Marscher14}. Optical to ultraviolet (UV) spectrum showed a rising shape in the higher frequency and the UV magnitude measured by {\it Swift}/UVOT was steady even during the flaring state, suggesting that thermal emission from accretion disk is dominant in that band.
\end{abstract}

\section{Introduction}

Narrow-Line Seyfert 1 galaxy (NLS1) is a sub-class of active galactic nuclei (AGN) and identified by the following criteria in optical emission lines: (1) narrow H$\beta$ line of FWHM(H$\beta$)$<$ 2000 km s$^{-1}$, (2) the flux ratio [O III] $\lambda$5007/H$\beta$ $<$ 3, (3) presence of emission lines from Fe II or high ionization lines such as [Fe VII] $\lambda$6087 and [Fe X] $\lambda$6375 \citep{Goodrich89}. The second and third criteria are to select Seyfert 1 galaxies and the first criterion is to choose objects of ``narrow-line" feature among the selected Seyfert 1 galaxies. The central black hole (BH) mass is usually derived as $M_{\rm BH}=v^2 R_{\rm BLR}/G$, where $v$ is the measured velocity dispersion, $R_{\rm BLR}$ is the radius of the broad line region (BLR), and $G$ is the gravitational constant, under the assumption that the BLR line-emitting materials are virialized. Here the BLR radius $R_{\rm BLR}$ can be derived from optical luminosity $L$ at e.g. 5100\AA\ or H$\beta$ based on a robust power-law relationship between $R_{\rm BLR}$ and $L$ established by reverberation mapping technique \citep[e.g.,][]{Peterson04, Kaspi05}. From these studies, it is widely believed that NLS1s have relatively small BH mass of $10^6-10^8 M_{\odot}$ \citep[e.g.,][]{Komossa06}, which is also supported by rapid X-ray variability \citep[e.g.,][]{Leighly99}.

Estimation of the central BH mass allows us to derive the accretion rate in Eddington unit based on the bolometric luminosity inferred from the optical luminosity at 5100 \AA\ and a bolometric correction factor of 9 ($L_{\rm bol}=9 \lambda L_{\lambda5100}$, see e.g., \citet{Warner04}). The resultant high value of the order of unity for $L_{\rm bol}/L_{\rm Edd}$ suggests that NLS1s have high accretion rate as a class \citep[e.g.,][]{Komossa06}. Combined with the small BH mass, NLS1 is considered to be the growing BH with high accretion rate, namely in the early phase of their evolution. It is essential to study this class in this regard, because this provides important information about how quasars are produced at the early Universe \citep[e.g.,][]{Mathur00,Grupe04}.

Recent \fermi\ discovery of MeV/GeV $\gamma$-ray emission from five radio-loud NLS1s confirmed the presence of relativistic jets aligned toward us \citep{Abdo09,Filippo12}, which were originally speculated by some authors mainly through the very long baseline interferometer (VLBI) radio observation \citep[e.g.,][]{Doi06,Doi07,Komossa06,Yuan08}. It is also found that spectral energy distributions (SEDs) of the {\it Fermi}-detected NLS1s are dominated by non-thermal emission across the whole frequency and consist of two broad humps, similarly to blazars \citep{Abdo09}. It is therefore believed that the low energy hump seen in the SED of the $\gamma$-ray loud NLS1s which extends from radio to optical or X-ray is synchrotron, while the higher-energy one is inverse Compton scattering of photons from synchrotron emission itself or external radiation field. Indeed, \citet{Itoh13} detected highly variable and significantly polarized optical emission from $\gamma$-ray loud NLS1 PMN~J0948+0022, which unambiguously indicates synchrotron origin of the optical emission, rather than the host galaxy or accretion disk, at least during high state for this object. However, we are still lacking knowledge of the optical emission mechanism for other $\gamma$-ray loud NLS1s.

It is known that NLS1s resides in spiral galaxies while blazars and radio galaxies which posses powerful relativistic jets are hosted by giant elliptical \citep[e.g.,][]{Marscher09}.
Therefore, recent VLBI and \fermi\ confirmation of presence of strong jets in some NLS1s is surprising and intriguing, because it suggests that such powerful relativistic jets can emerge also from spiral galaxies and seems to violate the well-known (though empirical) association between giant ellipticals and powerful jets. In this regards, detailed study of $\gamma$-ray loud NLS1s is important and expected to shed light on understanding the jet formation mechanism, in particular a physical link between the jet production and host galaxy environment.

\src\ is located at $z$=0.063 \citep{Zhou07}, nearest to us, and brightest in optical band among the known $\gamma$-ray loud NLS1s, and hence it is the best target and has been subject to multi-wavelength studies so far. For example, {\it Hubble Space Telescope} snapshot image revealed possible presence of one-armed spiral around the active nucleus \citep{Zhou07}, while \citet{Anton08} claimed from the 2.6~m Nordic Optical Telescope $B$- and $R$-band images that the structure seen around the nucleus is ring-like and that it may be thermal emission from star formation triggered by interactions and mergers.
A few separate methods for the estimation of the central BH mass for this object provided consistent result of $\left( 1-3 \right) \times 10^8 M_{\odot}$ \citep{Zhou07}. 

Here we show {\it Kanata} 2-year polarimetric monitoring results since 2012 September as well as optical and infrared dense follow-up results following the $\gamma$-ray flux enhancement in 2013 July detected by \fermi\footnote{Weekly \fermi\ information on flaring objects are available at \texttt{http://fermisky.blogspot.jp/}. See also \texttt{http://fermi.gsfc.nasa.gov/ssc/data/access/lat/msl\_lc/} for further reference.}. The latter consists of daily long-duration polarimetric and multi-band photometric observations performed by several small-diameter (0.5--2.0~m) telescopes all over Japan. We also present results of {\it Swift}-UVOT data taken at the corresponding period. We describe the observation in \S2 and the results are presented in \S3. We discuss interpretations and implications of our findings in \S4.

\section{Observation}

\src\ is included as one of the targets of blazar polarimetric monitoring program since 2012 September performed by the HOWPol instrument attached to the 1.5~m {\it Kanata} telescope located at the Higashi-Hiroshima Observatory, Japan \citep{Kawabata08}. This object was also observed as a Target of Opportunity (ToO) program of Optical and Infrared Synergetic Telescopes for Education and Research (\oister; see \citet{Itoh13oister} for detailed description of the available telescopes, instruments and observation frequencies). All the seven telescopes available are located all over Japan, which allow us to continuously observe the target without interruption by e.g., bad weather. This \oister\ ToO observation of \src\ was conducted from MJD 56485 to 56493 (corresponding to 2013 July 12 to 20) following a daily flux enhancement in MeV/GeV $\gamma$-ray band detected by \fermi. The result presented here were derived from $g^{\prime}$, $R_{\rm C}$, and $I_{\rm C}$ band photometric data taken by MITSuME (Multicolor Imaging Telescopes for Survey and Monstrous Explosions) detector systems attached to the 0.5~m telescopes located at Okayama Astrophysical Observatory (OAO) and Akeno Observatory in Japan \citep{Kotani05}. The same MITSuME system is also equipped with the 1.05\,m {\it Murikabushi} telescope located at the Ishigaki island. The $R_{\rm C}$- and $V$-band photometric as well as $R_{\rm C}$-band polarimetric data were taken by the 1.5~m {\it Kanata} telescope. 

In addition to these ground-based telescope materials, we analyzed the data taken by the UV and Optical Telescope (UVOT) onboard {\it Swift} satellite. {\it Swift}/UVOT provides optical and ultraviolet (UV) data which covers the wavelength range of 170--650~nm by utilizing 6 filters ($UVW2$, $UVM2$, $UVW1$, $U$, $B$, $V$ bands). The UVOT data analyzed here are taken on MJDs 56485, 56489, 56491 and 56492 and they are downloaded from {\it Swift} Data Center\footnote{\texttt{http://swift.gsfc.nasa.gov/sdc/}}. The {\it Swift} observation ID are 00036533038, 00036533039, 00036533040, 00036533041 and 00036533042.

\subsection{Photometry}

We reduced the data following the standard procedure of CCD photometry. We performed differential aperture photometry using a comparison star located at R.A. = $03^h 24^m 39.6^s$, Decl. = $+34\arcdeg\ 11\arcmin\ 29.8\arcsec.1$ (J2000). The magnitudes of $g^{\prime}$=14.314 mag, $V$=13.510 mag, and $R_{\rm C}$=13.496 mag are taken from SDSS database \citep{SDSS07}. The Galactic extinction for the direction of \src\ of $A_{\rm V}$=0.680 was taken from NED database\footnote{\texttt{http://ned.ipac.caltech.edu/}} \citep{Schlafly11}, and then applied to derive the intrinsic magnitudes. The systematic difference of the $R_{\rm C}$-band magnitude among the observatories is calibrated within $\Delta R_{\rm{C}}\sim 0.05$ mag, which was added to the photometric error.

A photometric error consists of statistical and systematic ones. The $R_{\rm C}$-band magnitudes shown in Fig.~\ref{fig:longlc} and \ref{fig:shortlc} were measured at various observatories and so we included systematic error of $\sim$0.05 mag in each data point, as is described above. On the other hand, for {\it Ishigaki} data shown in Fig.~\ref{fig:ishigaki}, we do not need to take into account the systematic uncertainty among observatories and hence only statistical errors are included. The statistical error comes from uncertainty of photon numbers measured from a circular region centered at the source position. This is derived from \texttt{APPHOT} task in \texttt{PYRAF} package. The statistical photometric errors of {\it Ishigaki} 3-band data are mostly 0.01--0.02~mag, and they are used for calculation of excess variance (see \S3 for details). Here, we checked a systematic photometric uncertainty caused by e.g., bad weather and temporal variation of a comparison star by making a light curve of a comparison star during the same time interval. We fit the light curve by a constant value and derived a variance in each band. As a result, we obtained that the $g^{\prime}$-, $R_{\rm C}$-, and $I_{\rm C}$-band standard deviations (namely, systematic uncertainties) are 0.004, 0.002, and 0.003 mag, respectively, which are all smaller than the statistical photometric errors of 0.01--0.02 mag derived by APPHOT task. In addition, it is known that host galaxy of \src\ is present in the {\it Hubble Space Telescope} image and it is as bright as the central core \citep{Zhou07}. This may cause an additional photometric error for the flux measurement of the central AGN. According to \citet{Cellone00}, when magnitudes of the central AGN and the host spiral galaxy are the same ($V$=16~mag), photometric error of the central AGN is less than 0.01~mag if aperture radius for photometry is larger than 3.3$^{\prime\prime}$ and seeing change during exposure is less than 1$^{\prime\prime}$. In our photometry, the aperture radius is set to 10$^{\prime\prime}$. We also checked the seeing during the $Ishigaki$ 2-hour exposure and found that it is mostly between 2.75$^{\prime\prime}$ and 3.75$^{\prime\prime}$. Hence, we can utilize the result described in \citet{Cellone00} (see their Figures~8 and 9 and Table~2) and robustly claim that photometric error of the central AGN caused by host galaxy contamination is less than 0.01~mag and negligibly small. The same conclusion is also derived by \citet{Paliya13} (see Section 4.1 of his/her paper).

Data reduction of the {\it Swift}/UVOT data was also performed under the standard procedure of CCD photometry using \verb|APPHOT| packaged in \verb|PYRAF|. For each filter system, the conversion factor is well calibrated by the UVOT team\footnote{{\tiny \texttt{http://heasarc.gsfc.nasa.gov/docs/heasarc/caldb/swift/docs/uvot/index.html}}}. We extracted source count from a circular region of 5'' radius centered at the source position and converted them to fluxes using the standard zero points \citep{2008MNRAS.383..627P}. Then, the Galactic extinction was corrected (e.g., $A$(UVW2)=0.65 \citealt{1998ApJ...500..525S}).

\subsection{Polarimetry}

We performed continuous photo-polarimetric observations for \src\ using the HOWPol with durations of several hours on MJDs 56486, 56488, 56489 56491 and 56492. We obtained both the Stokes parameters $Q$ and $U$ in one-shot 100 s exposure thanks to the double Wollaston prism installed inside the HOWPol. But since these continuous observations were often interrupted by bad weather, the data qualities were not enough to search for any significant intra-night (hour-scale or minute-scale) variation. We therefore averaged all the data taken on the same night and derived the daily polarization degree (PD) and angle (PA) to find out their possible temporal variations. Here we carefully checked the systematics of the PD and PA in our HOWPol measurements and confirmed that the systematic errors of the PD and PA are less than 0.5\% and $2^{\circ}$, respectively, from observations of strongly-polarized and unpolarized stars. We further checked the systematic errors of the Stokes parameters by using a comparison star within the field of view of \src. As a result, we obtained the systematic errors of $\Delta Q\sim 0.003$ and $\Delta U \sim 0.003$ and these errors were added to the statistical ones.

\section{Result}

\begin{figure}
 \begin{center}
  \includegraphics[width=9cm]{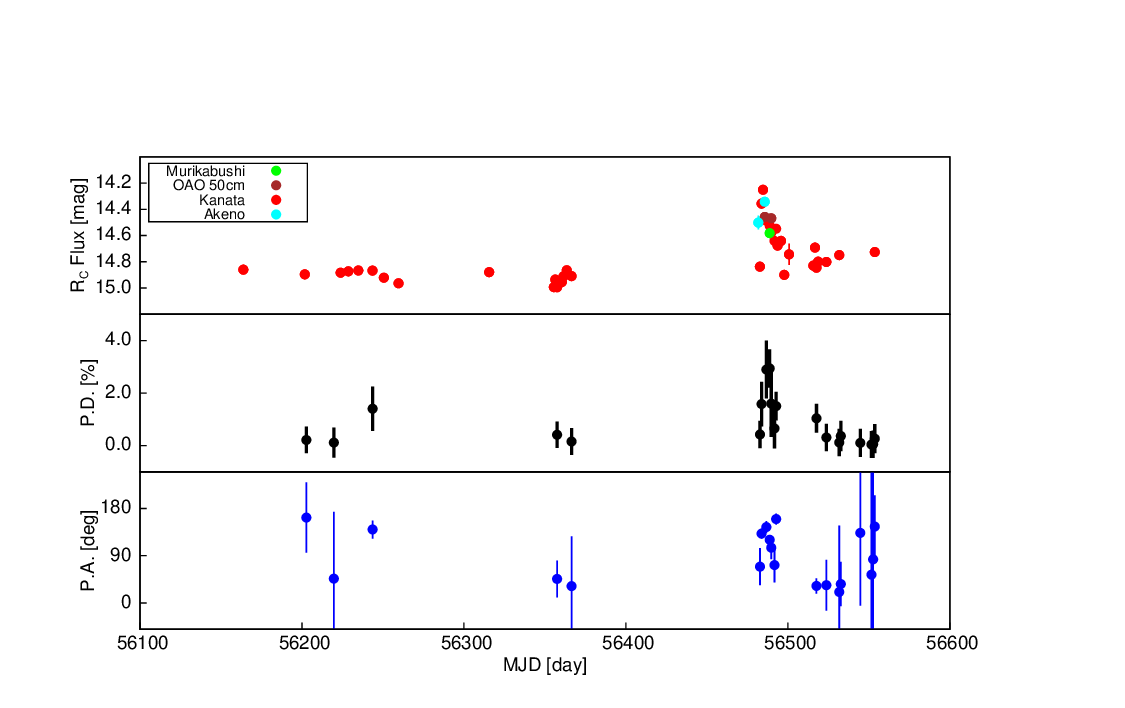} 
 \end{center}
\caption{Long-term variations of daily-binned flux, polarization degree, and polarization angle of \src, all of which are measured in $R_{\rm C}$-band. A close-up view between MJD 56480 and 56530 are shown in Fig.~\ref{fig:shortlc}. Different colors in the upper panel represent the magnitudes measured by different telescopes.}\label{fig:longlc}
\end{figure}

\begin{figure}
 \begin{center}
  \includegraphics[width=9cm]{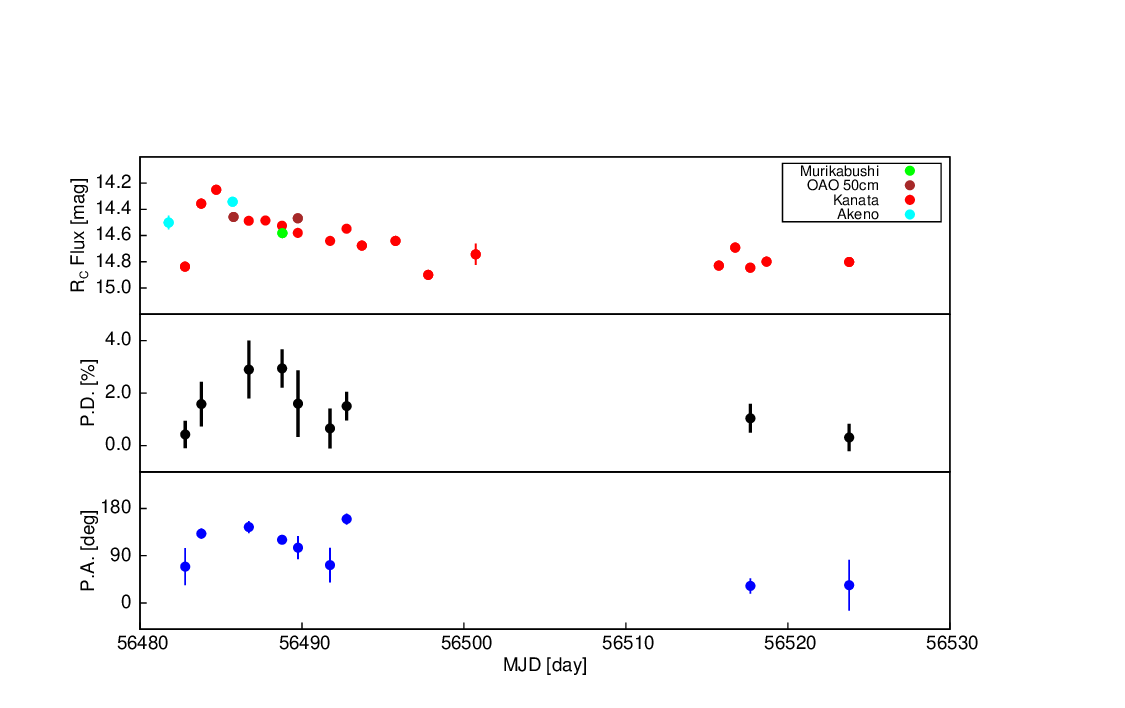} 
 \end{center}
\caption{Same as Fig.~\ref{fig:longlc} but the plotted time range is limited to MJD 56480 to 56530.}\label{fig:shortlc}
\end{figure}

Fig.~\ref{fig:longlc} shows the daily-binned $R_{\rm C}$-band light curve of \src, together with the PD and PA measured by \kanata/HOWPol. The optical flux was almost constant and probably in quiescence until clear brightening on MJD~56481 and subsequent $\sim$20 days, although the monitoring frequency was not very dense. During the low state, the $R_{\rm C}$-band emission is almost unpolarized and the PD was 0--1\% (at most $1.4\pm0.8$\% on MJD~56243), consistent with past measurements by \citet{Ikejiri11} and \citet{Eggen11}, while the PAs have large errors because of the very low PD. In contrast, we detected clear increase of the PD up to 2.9$\pm$0.5\% during the high state which starts on MJD~56482. To see the temporal variations in detail, we show in Fig.~\ref{fig:shortlc} a close-up view of Fig.~\ref{fig:longlc} during the flaring state. Although complete rising profile was not obtained, the $R_{\rm C}$-band light curve showed rapid dimming on MJD~56482, peak on MJD~56484, and subsequent gradual decay. The PD reached a maximum on MJDs~56486 and 56488, several days after the peak of the $R_{\rm C}$-band flux. During the PD increase, we were able to precisely measure the PA and found that it was scattered somewhat but roughly constant at 105--145 deg around the PD peak.

\begin{figure}
 \begin{center}
  \includegraphics[width=9cm]{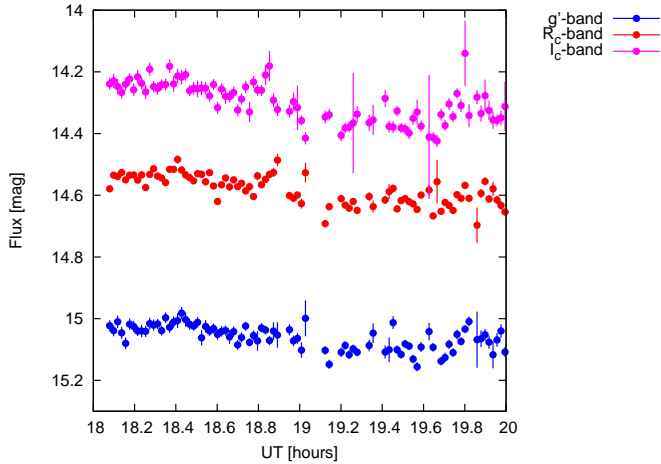} 
 \end{center}
\caption{2-hour continuous light curves of \src\ observed on MJD 56488 by the 1.05~m {\it Murikabushi} telescope located at Ishigaki island in Japan. The instrument attached to the {\it Murikabushi} telescope enables us to obtain the light curves in $g^{\prime}$, $R$, and $I$ bands simultaneously.}\label{fig:ishigaki}
\end{figure}

\begin{figure}
 \begin{center}
  \includegraphics[width=8cm]{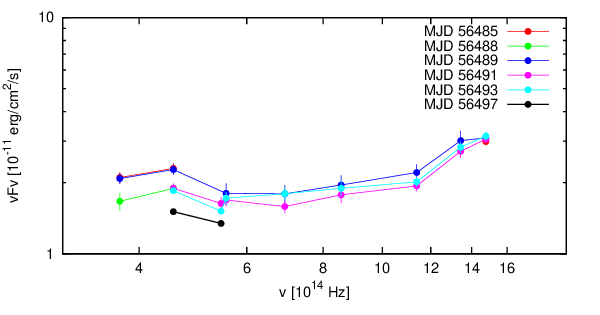} 
 \end{center}
\caption{Daily evolution of the optical-ultraviolet spectral energy distribution of \src\ during the high state. Note that the SED data points on e.g. MJD~56493 are obtained by combining the optical and {\it Swift/UVOT} data
taken on MJD~56492.5--56493.5 based on the actual observation times. }\label{fig:sed}
\end{figure}

Using the 1.05~m {\it Murikabushi} telescope, we obtained continuous (2-hour) and simultaneous $g^{\prime}$-, $R_{\rm C}$-, and $I_{\rm C}$-band light curves of \src\ on MJD~56488, which are shown in Fig.~\ref{fig:ishigaki}. All the light curves showed a common trend during the exposure, i.e., gradual dimming (18--19.1 UT) followed by constant flux (19.1--19.6 UT), and then small brightening peaking around 19.8 UT. Small-amplitude fluctuations superposed on the trend are also visible. We also found from Fig.~\ref{fig:ishigaki} that the flux variation in $g^{\prime}$-band seems to be smaller than that of $I_{\rm C}$-band, and hence in order to quantify the amplitude of variability in each band we calculated the normalized ``excess variance" ($\sigma^2_{\rm rms}$, see \citet{Nandra97} for the calculation method) for the $g^{\prime}$-, $R_{\rm C}$-, and $I_{\rm C}$-band light curves. Here we removed several (3 at most) data points which have large errors of the order of 0.1~mag in each light curve. The results are tabulated in Table~\ref{tab}. The higher frequency light curve showed lower variability amplitude, implying that variable synchrotron emission is more contaminated by thermal emission from accretion disk in higher frequency. Here we note that synchrotron emission from the jet is highly variable due to relativistic beaming effect, while disk emission can be safely assumed to be steady at least during the 2-hour exposure, as is verified from the {\it Swift}/UVOT measurement of constant flux in $UVW2$ band during high state. To check this possibility as well as spectral evolution, we constructed daily optical-UV SEDs and they are displayed in Fig.~\ref{fig:sed}. The higher frequency portion of the SEDs clearly showed a rising shape, indicating again the presence of disk emission. No significant flux variation in the highest frequency at $1.5 \times 10^{15}$~Hz ($UVW2$ band) also supports the disk origin. 

\section{Discussion}

The broad-band SED of \src\ is reasonably modeled by one-zone synchrotron plus inverse compton scenario, as is the case of blazars \citep{Abdo09}. The optical emission was therefore believed to be synchrotron radiation, but the very low PDs of 0--1\% by past measurements \citep{Ikejiri11,Eggen11} challenged the synchrotron origin to some extent. One of the new findings presented here is the first clear detection of variable polarized emission from this object, and the observed PD was moderate ($\sim 3\%$ at most). This positive detection undoubtedly indicates the synchrotron origin for the optical emission. The $R_{\rm C}$-band light curve during high state showed gradual decay of typical e-folding timescale of several days and if we interpret this as synchrotron cooling time of relativistic electrons emitting optical photons, we can roughly estimate the magnetic field strength at the emission region as $B \simeq 0.25 \left(t_{\rm decay}/3 \ {\rm days} \right)^{-2/3} \left( E_{\rm obs}/1 \ {\rm eV} \right)^{-1/3} \left( \delta/10 \right)^{-1/3}$ G, where $t_{\rm decay}$ and $E_{\rm obs}$ are measured in the observer's frame and $\delta$ is a relativistic beaming factor and 10 is assumed here as usual for blazars. The derived magnetic field is lower than that inferred from SED fitting for the multi-wavelength data \citep{Abdo09}, where long-term non-simultaneous data were used.

\begin{table}
  \caption{Excess variance $\sigma^2_{\rm rms}$ calculated from the {\it Ishigaki} simultaneous 3-band light curves shown in Fig~\ref{fig:ishigaki}.}\label{tab}
  \begin{center}
    \begin{tabular}{cc}
      \hline
      Band & Value \\
      \hline
      $g^{\prime}$ & $\left( 4.3\pm1.6 \right) \times 10^{-6}$ \\
      $R_{\rm C}$ & $\left( 9.0\pm2.5 \right) \times 10^{-6}$ \\
      $I_{\rm C}$ & $\left( 14.5\pm3.2 \right) \times 10^{-6}$ \\
      \hline
    \end{tabular}
  \end{center}
\end{table}

The PAs measured reliably during the PD enhancement were scattered at 105--145 degree but roughly constant, and did not show significant temporal variation such as rotation/swing seen in some blazars \citep[e.g.,][]{Marscher08,Abdo10}. Here we note a similarity that the PAs were always almost constant whenever polarized flux flares were observed by {\it Kanata} telescope. See Figure~1 of \citet{Itoh13oister} for the case of a famous flat spectrum radio quasar CTA~102 , and Figure~2 of \citet{Itoh13} for $\gamma$-ray loud NLS1 PMN~J0948+0022. In light of the turbulent multi-zone model recently proposed by \citet{Marscher14}, this can be naturally understood by brightening of a single highly ordered magnetic field cell (or emission region) among multiple radiating cells. In contrast, the $R_{\rm C}$-band flux enhancement with low PD at e.g. MJD 56491 would be caused by superposition of brightening of multiple cells, because the random orientation of the magnetic field in each cell would cancel the polarization in total and as a result the PD was observed to be low. The small-amplitude intra-night rapid variability observed during the high state (see Fig.~\ref{fig:ishigaki} and also \citet{Paliya13} for past detection from the same object) would also be explained by twinkling multiple small cells within the framework of this multi-zone scenario.

Parsec-scale structure of the jet has been resolved by VLBI observation and the position angles of the innermost components (i.e., jet direction at the inner region accessible by VLBI) are measured as 111--131 deg \citep[see Table~5 of][]{Wajima14}. Hence, the optical PAs of 105--145 deg observed during the high state roughly correspond to the jet direction under a reasonable assumption that the jet direction at sub-pc scale is the same as the that of pc-scale resolved by VLBI. This implies that the magnetic field direction at the optical emission region is roughly perpendicular to the jet axis, because the magnetic field direction is assumed to be perpendicular to the observed PA. The inferred magnetic field direction transverse to the jet axis may indicate the presence of helical/toroidal magnetic field at the emission site. But non-detection of PA rotation seems to contradict with this hypothesis. An alternative and more plausible option would be ``shock-in-jet" scenario where magnetic field is compressed and aligned to the direction perpendicular to the jet axis by a transverse shock propagating downstream \citep[e.g.,][]{Laing80,Hughes85,Hagen08}. We speculate that the ordered magnetic field within the single cell responsible for the PD enhancement would be formed by well-known internal shock, i.e., blobs (or shells) intermittently emitted from the central BH collide, shock is formed there, and particle acceleration and ordering of the magnetic field to the direction transverse to the jet take place. The colliding blobs may merge into single one which corresponds to the cell described above. However, it would be difficult to discriminate the helical/toroidal magnetic field and ``shock-in-jet" scenarios based only on the optical results. In this regard, measurement of Faraday rotation measure gradient across the jet provides direct evidence for the presence of helical/toroidal magnetic field \citep[e.g.,][]{Asada02,Gabuzda04}, but this technique does not allow us to investigate the magnetic field structure inside the radio core.

From the analysis of the {\it Ishigaki} 3-band simultaneous and continuous 2-hour light curves, we found a clear trend of smaller variability amplitude for the higher frequency light curve. This can be naturally interpreted that steady disk emission is dominant over variable synchrotron emission in higher frequency. We note that the disk emission is also variable in principle, but at least during the high state analyzed here we did not find any significant flux variation in the {\it Swift}/UVOT data in $UVW2$ band, where the disk emission is most dominant among all the other band data. Hence we can safely assume the constant UV flux during the {\it Ishigaki} 2-hour exposure and this validates the above interpretation. We also found that the UV spectrum of the source showed a rising shape. All of these findings presented here indicates that UV radiation of \src\ is dominated by thermal emission from accretion disk.

In summary, we detected optical polarized emission from \src\ during high state, providing clear evidence of synchrotron origin for the optical emission. The PAs measured during the high state was roughly aligned to the jet axis. This indicates that the magnetic field direction at the optical emission region was roughly transverse to the jet axis, which implies either that helical/toroidal magnetic field is there or magnetic field is compressed by transverse shock propagating down the jet. The UV spectrum of \src\ showed a rising shape and did not show significant variability during the high state. This would suggest that the UV radiation of the source is dominated by thermal emission from accretion disk. This is also supported by our finding that higher frequency light curve showed less variability amplitude during the {\it Ishigaki} 2-hour continuous exposure.

The central BH mass of radio-loud NLS1s is still controversial. Radio-loud NLS1s may have heavy BHs above 10$^8$ $M_{\odot}$, comparable to typical blazars, under the assumption of standard disk spectra for these objects \citep{Calderone13}. If so, Eddington ratio reduces to 0.04--0.2 and are not extreme. Here we note that our conclusions presented in this paper are not affected by uncertainty of the central black hole mass estimation.

Finally we mention that new optical-infrared instrument HONIR (Hiroshima Optical and Near-InfraRed camera, see \citet{HONIR12} and \citet{HONIR14} for detailed description) has been attached to {\it Kanata} telescope and HONIR observation in polarization mode has started to be operated since 2014 January. Multi-band simultaneous polarimetry is now feasible and HONIR observation of the optical and infrared polarization, together with {\it OISTER} facilities, will bring important information on blazar science in near future.

\bigskip
We appreciate the referee's critical reading and valuable comments.
R.I., Y.T.T. and H.A. led this project. R.I. has led the daily {\it Kanata} blazar monitoring program, and R.I. and H.A. led the {\it OISTER} observation after
the ToO trigger by Y.T.T. R.I. led the overall data analysis. Y.T.T. led the interpretation of the results as well as drafting of the manuscript.

%5\begin{thebibliography}{}
% Journals(e.g. A\&A,ApJ,AJ,NMRAS,PASP ...)
% Authors, Year, Journal, Vol#, Page#
% Journal Title Abbreviation >> http://www.asj.or.jp/pasj/Jabb.html
%\bibitem[Aauthor et al.(2001)]{key-1}
 % Aauthor, A., Bauthor, B., Cauthor, C.\ 2001, PASJ, vol, page
% Books
%\bibitem[Aauthor \& Author(2001a)]{key-2}
 % Aauthor, A., Author, B.\ 2001, Name of Book(Publisher, Tokyo) ch.0
% Books
%\bibitem[Aauthor \& Bauthor(2001b)]{key-3}
 % Aauthor, A., Bauthor, B.\ 2001, Name of Book(Publisher, Tokyo) page0
%cc
% Editorial Books
%\bibitem[Dauthor(2001)]{key-n}
 % Dauthor A.~A.\ 2001, in Name of Book,
   %ed.\  D.~Editor (Publisher, Tokyo) page0
%\end{thebibliography}

\end{document}